\documentclass[preprint,prd,aps,pdftex,showpacs,preprintnumbers,amsmath,amssymb]{revtex4-1}
\usepackage{graphics,epsfig,subfigure}
\usepackage{diagbox}
\usepackage[usenames]{color}
\usepackage{color}
\usepackage{graphicx}
\usepackage{amsfonts}
\usepackage{indentfirst}
\usepackage{booktabs}
\usepackage{hyperref}
\hypersetup{hypertex=ture,backref=true,colorlinks=ture,linkcolor=blue,anchorcolor=blue,citecolor=blue}
\usepackage{float}
\usepackage{multirow}

\begin{document}

\title{Frozen Hayward-boson  stars}

\author{Yuan Yue$^{1}$}
\author{Yong-Qiang Wang$^{2}$}\email{yqwang@lzu.edu.cn, corresponding author}
\affiliation{$^{1}$College of Mathematics and Computer Science, Northwest Minzu University, Lanzhou, 730030, China\\
$^{2}$School of Physical Science and Technology, Lanzhou University, Lanzhou 730000, China}

\begin{abstract}
Recently, the model of  the Einstein-Bardeen theory minimally coupled to a complex, massive, free scalar field was investigated in arXiv:2305.19057. 
The introduction of a scalar field disrupts the formation of an event horizon, leaving only a type of solution referred to as a Bardeen-boson star. When the magnetic charge $q$ exceeds a certain critical value, the frozen Bardeen-boson  star can be obtained with $\omega \rightarrow 0$.  In this paper,  we extend  to  the investigation of Einstein-Hayward-scalar model, and obtain the solution of frozen Hayward-boson  star, including the ground and excited states.
Furthermore,   under the same
parameters, it is interesting to observe that both the ground state and the excited states frozen stars have the same critical
horizon and mass.
\end{abstract}

\maketitle

\section{Introduction}
As a pivotal prediction of general relativity, black holes have consistently been a focal point of research. However, the early  discovered exact solutions, such as Schwarzschild and RN black holes, manifest singularities with infinite curvature at their centers. This development has spurred the formulation of singularity theorems by R. Penrose and S. Hawking \cite{Penrose:1964wq,Hawking:1966sx,Hawking:1970zqf}, asserting the universal and inevitable existence of singularities when considering  the validity of strong energy conditions 
and  the   globally hyperbolic spacetime  in Einstein’s general relativity. Moreover,
the singularity theorems prove that a singularity will always form once an event horizon forms, which  implies that the emergence of black holes should be viewed as an inherent occurrence within the evolutionary trajectory of the universe.
Nevertheless, in reality, given that matter cannot be compressed infinitely \cite{Einstein:1939ms}, black holes should not harbor singularities. Singularities are flawed and non-physical mathematical outcomes.
The prevalent consensus holds that general relativity serves as a low-energy effective framework for quantum gravity, with the anticipation that a comprehensive quantum gravity theory would impede the occurrence of singularities.  However, If we abandon the constraints imposed by the strong energy condition,  the singularities within a black hole's interior can be eliminated by employing a certain special forms of matter field.
 The initial regular black hole solution to Einstein's equations was derived by J. Bardeen \cite{Bardeen1}. Bardeen's solution arises from solving Einstein's equations in the presence of a nonlinear electromagnetic field \cite{Ayon-Beato:1998hmi,Ayon-Beato:2000mjt}. Currently, there is a wealth of literature on the study of regular black holes, garnering growing attention and undergoing continuous development (see review \cite{Lan:2023cvz}).

Recently,  the model of the coupling of an  complex scalar field to   Bardeen spacetime is investageted in \cite{Wang:2023tdz}, This model comprises two special solution cases: one is the Bardeen spacetime solution, and the other is the boson star solution. The concept of boson stars originated from J. Wheeler's geon theory \cite{Wheeler:1955zz,Power:1957zz}. In 1960's, the theoretical framework of boson stars was established using a scalar field \cite{Kaup:1968zz,PhysRev.187.1767}. See Refs. \cite{Schunck:2003kk,Liebling:2012fv} for a review. If both a nonlinear Maxwell field and a scalar field exist simultaneously, the result is a solution known as the Bardeen-boson star. Surprisingly, there is no black hole solution with an event horizon (EH) in this context.
When the magnetic charge $q$ exceeds a certain critical value, the frozen Bardeen-boson  star can be obtained with $\omega \rightarrow 0$. In this scenario, the scalar field converges at the critical horizon, experiencing rapid decay beyond this threshold.
Inside the  critical horizon of  star, 
the $g_{tt}$ is nearly zero.
From the vantage point of an observer at infinity, one can interpret these solutions of  stars as analogous to an extremal black hole. These above behavior exhibited by these types of solutions is indicative of the characteristics of a frozen star.
A frozen star is a theoretical concept that originated from the problem of black hole  formation from
gravitational collapse studied by J. Oppenheimer and H. Snyder \cite{Oppenheimer:1939ue}, and  
named by Y. Zel'dovich and I. Novikov ~\cite{zeldovichbookorpaper}. The term ``frozen" is used because, when observed from a distant perspective, the collapse of an ultra-compact object appears to occur over an extended period, creating that the star is frozen at their own gravitational radius \cite{Ruffini:1971bza}.

This paper extends the investigation of the frozen Bardeen-boson  star in  \cite{Wang:2023tdz} to the frozen Hayward-boson  star case, exploring the Einstein-Klein-Gordon theory coupled to a nonlinear electrodynamics proposed by S. Hayward \cite{Hayward:2005gi}. 
We find out the  ground and excited  state solutions of the Hayward-boson star, where there is no event horizon, and frozen star solutions emerge under specific conditions.
Furthermore, we explore the range of existence of these excited state solutions and analyze
their associated physical properties.

The paper is organized as follows.  In Section. \ref{model}, we provides an introduction to the model of Einstein-Hayward theory coupled to a free complex scalar field.
 In Section \ref{sec3}, we present numerical results of Hayward-boson star and a comprehensive analysis of their physical properties. The conclusion and discussion
are given in the last section.

\section{The Model}\label{model}
In this section, we aim to provide a concise introduction to the theoretical framework encompassing the Einstein-nonlinear electrodynamics model, coupled with a free complex scalar field. The expression for the bulk action is delineated as follows:
\begin{equation}\label{action}
  S=\int\sqrt{-g}d^4x\left(\frac{R}{4}+\mathcal{L}^{(1)}+\mathcal{L}^{(2)}\right),
\end{equation}
with
\begin{eqnarray}
\mathcal{L}^{(1)} &= &- \frac{ 3}{ 2 s } \frac{ (2 q^2 {\cal F})^{3/2}}{\left(  1 + ( 2 q^2 {\cal F})^{3/4}\right)^2} ,\\
\mathcal{L}^{(2)} &= & -\nabla_a\psi^*\nabla^a\psi  - \mu^2\psi\psi^*,
\end{eqnarray}
where $R$ denotes the scalar curvature, and $\mathcal{L}^{(1)}$ is a function dependent on ${\cal F} = \frac{1}{4}F_{ab} F^{ab}$ involving the electromagnetic field strength $F_{ab} = \partial_{a} A_{ b} - \partial_{b} A_{ a}$, with $A$ representing the electromagnetic field. The complex scalar field $\psi$ is introduced, and the constants $q$, $s$, and $\mu$ serve as three independent parameters, where $q$ signifies the magnetic charge, and $\mu$ denotes the scalar field mass. By variating the action (1) with respect to the metric, electromagnetic field, and scalar field, we obtain the following equations of motion
\begin{eqnarray} \label{eq:EKG1}
R_{ab}-\frac{1}{2}g_{ab}R-2 (T^{(1)}_{ab}+T^{(2)}_{ab})&=&0 \ ,  \nonumber\\
\bigtriangledown_{a} \left(\frac{ \partial {\cal L}^{(1)}}{ \partial {\cal F}}  F^{a b}\right) &=& 0,    \\
\Box\psi-\mu^2\psi &=& 0, \nonumber
\end{eqnarray}
with
\begin{equation}
T^{(1)}_{ab} =- \frac{ \partial {\cal L}^{(1)}}{ \partial {\cal F}} F_{a c} F_{ b }^{\;\;c} + g_{ab} {\cal L}^{(1)},
\end{equation}
\begin{equation}
T^{(2)}_{ab} = \partial_a\psi^*\partial_b\psi + \partial_b\psi^*\partial_a\psi - g_{ab}\left[\frac{1}{2}g^{ab}\left(\partial_a\psi^*\partial_b\psi + \partial_b\psi^*\partial_a\psi\right) + \mu^2\psi^*\psi\right]\,.
\end{equation}

In accordance with Noether's theorem, the action's invariance concerning a complex scalar field under the $U(1)$ transformation, denoted as $\psi\rightarrow e^{i\alpha}\psi$ where $\alpha$ remains a constant, yields a conserved current linked to this complex scalar field
\begin{equation}\label{equ8}
	J^{a} = -i\left(\psi^*\partial^a\psi - \psi\partial^a\psi^*\right).
\end{equation}
Through the integration of the timelike component of the above conserved current over a spacelike hypersurface denoted as $\varSigma$, the Noether charge pertaining to the complex scalar field can be ascertained as
\begin{equation}\label{equ9}
	Q = \frac{1}{4\pi}\int_{\varSigma}J^t .
\end{equation}

We contemplate a generic static spherically symmetric solution and employ the following ansatz:
\begin{equation}\label{equ10}
	ds^2 = -N(r)\sigma^2(r)dt^2 + \frac{dr^2}{N(r)} + r^2\left(d\theta^2 + \sin^2\theta d\varphi^2\right).
\end{equation}
Herein, the functions $N(r)$ and $\sigma(r)$  depend exclusively  on the radial variable $r$. Additionally, we employ the subsequent ansatzes for the electromagnetic field and the scalar field
\begin{equation}\label{equ11}
   A= q \cos(\theta)d\varphi,\;\;\; \psi = \phi(r)e^{-i\omega t},
\end{equation}
Thus,	with the ansatz of electromagnetic field in Eq. (\ref{equ11}),
the magnetic field is given by
\begin{equation}
F_{\theta \varphi } =  2 q \sin \theta,
\end{equation}
we can determine the magnetic charge of the magnetic monopole
\begin{equation}\label{equ18}
	\frac{1}{4\pi}\oint_{S^\infty}d A =q  .
\end{equation}
Besides, the Noether charges obtained from  Eq.~(\ref{equ9}) are written as
\begin{equation}\label{equ18}
	Q = 2\int_0^\infty r^2\frac{\omega\phi^2}{N~\sigma}dr\, .
\end{equation}

By substituting the above ansatzes (\ref{equ10}) and (\ref{equ11}) into the field equations (\ref{eq:EKG1}), we derive the following equations governing $\phi(r)$, $N(r)$, and $\sigma(r)$:
\begin{eqnarray}
	 \phi^{\prime\prime}+\left(\frac{2}{r} + \frac{N^\prime}{N} + \frac{\sigma^\prime}{\sigma}\right)\phi^\prime + \left(\frac{\omega^2}{N \sigma^2} - \mu^2\right)\frac{\phi}{N} &=& 0\, ,\label{ode11}\\
 N'+2 \mu^2 r \phi^2+\frac{2 r \text{$\omega $}^2 \phi^2}{N \sigma^2}+2 r N \phi'^2+\frac{N}{r}+\frac{3 q^6 r}{s \left(q^3+r^3\right)^{2}}-\frac{1}{r}&=&0, \label{ode12}
 \\
  	\frac{\sigma^\prime}{\sigma} - 2r\left(\phi^{\prime2} + \frac{\omega^2\phi^2}{N^2 \sigma^2}\right)&=&0.  \label{ode13}
\end{eqnarray}

In the pursuit of solving the above set of ordinary differential equations, it becomes crucial to specify  suitable boundary conditions for each unknown function. Given the asymptotically flat characteristics of the solutions, the metric functions $N(r)$ and $\sigma(r)$ are required to satisfy the ensuing boundary conditions:
\begin{equation}\label{equ19}
N(0) = 1, \qquad \sigma(0) = \sigma_0, \qquad N(\infty) = 1-\frac{2 M}{r}, \qquad \sigma(\infty) = 1.
\end{equation}
For the  constants, namely $\sigma_0$ and the mass $M$ of the solution, their values can be deduced by solving the system of differential equations. Furthermore, specific boundary conditions are mandated for the complex scalar field:

\begin{equation}\label{equ20}
\phi(\infty) = 0,\;\;\;\left. \frac{d\phi(r)}{dr}\right|_{r = 0} = 0.
\end{equation}

These equations of motion  (\ref{ode11}), (\ref{ode12}) and  (\ref{ode13}) possesses solutions for two distinct special cases.
 Firstly, when $q = 0$, the action (\ref{action}) reduces to Einstein-scalar theory. The corresponding solution, capable of depicting an  spherically soliton, is called as  boson star.
In the case of a vanishing complex scalar field with $q \neq 0$, the model aligns with Bardeen model, and the associated solution, describing both  spherically solitons and black holes, manifests as the Bardeen spacetime. The metric is expressed in the following form
\begin{equation}
ds^2 = -f(r) dt^2 + f(r)^{-1} dr^2 + r^2 ( d \theta^2 + \sin^2(\theta) d \varphi^2),
\end{equation}
with
\begin{equation}
f(r) = 1 - \frac{2M r^2}{ ( r^3 + q^3 ) },
\end{equation}
where $M = \frac{q^3}{2s}$. The function $f(r)$ demonstrates a local minimum value at $r = 2^{1/3} q$. In case where $q < \frac{\sqrt{3s}}{2^{1/3}}$, the solutions without event horizons are present. When $q = \frac{\sqrt{3s}}{2^{1/3}}$, degenerate horizons are observed. Conversely, for $q > \frac{\sqrt{3s}}{2^{1/3}}$, there are two distinct horizons.

\section{Numerical results}\label{sec3}
In this section, 
we will solve the above coupled Eqs. (\ref{ode11}), (\ref{ode12}), and (\ref{ode13}) with boundary conditions (\ref{equ19}) and (\ref{equ20}) numerically.  It
is convenient to change the radial coordinate $r$ to a new radial coordinate $x=\frac{r}{1+r}$. 
 Consequently, the boundaries of the computed region are set at $x=0$ and $x=1$, respectively. Employing a common pseudo-spectral collocation method \cite{Trefethen}, we integrate the nonlinear ordinary differential equations utilizing a Newton–Raphson technique. All functions are represented as series expansions of Chebyshev polynomials in 
$x$, and the integration region spans $0\geq x\geq 1$. The estimated relative error of the numerical solutions in our study remains below $10^{-6}$.

\begin{figure}
  \includegraphics[width=8.2cm]{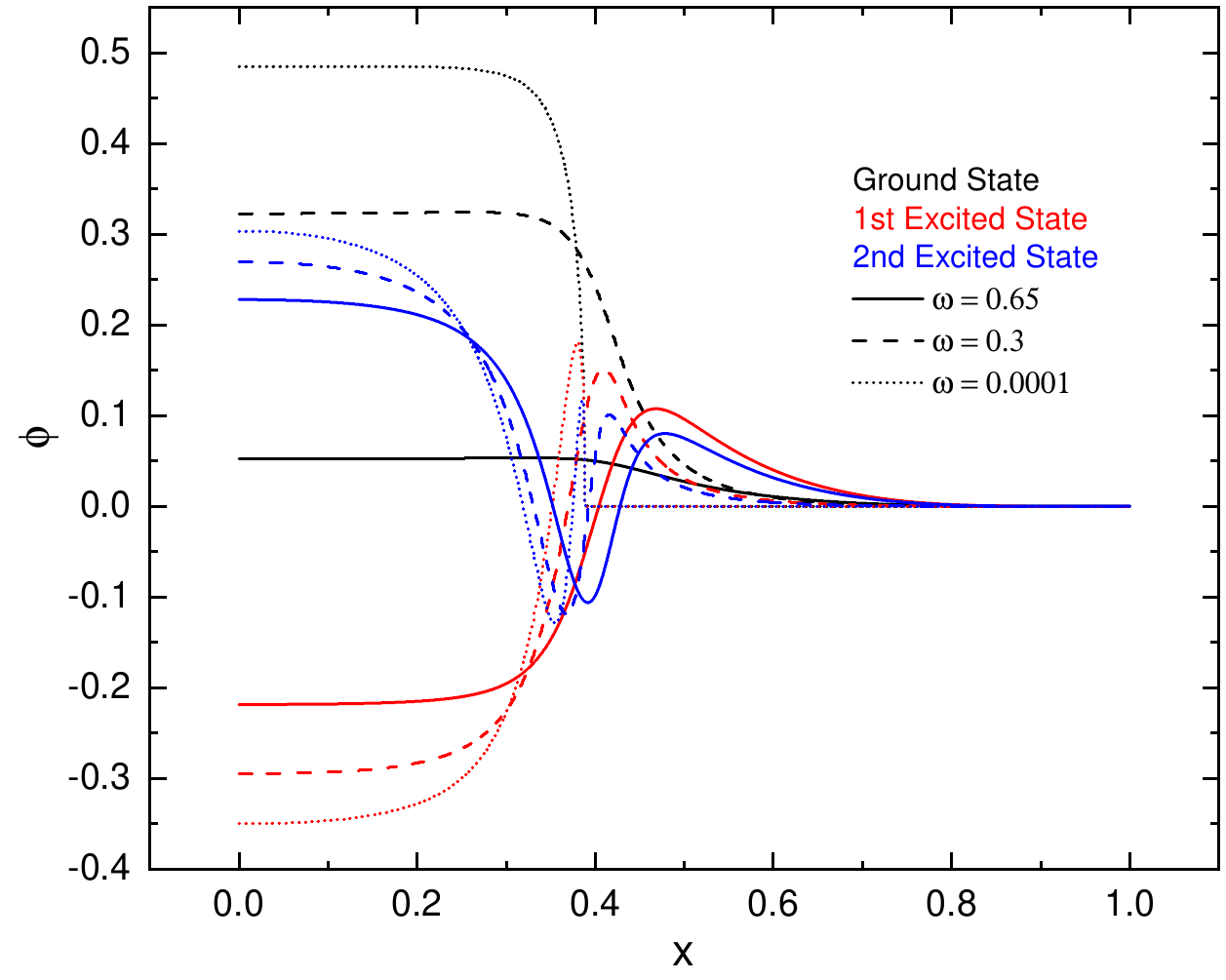}
 \includegraphics[width=8.1cm]{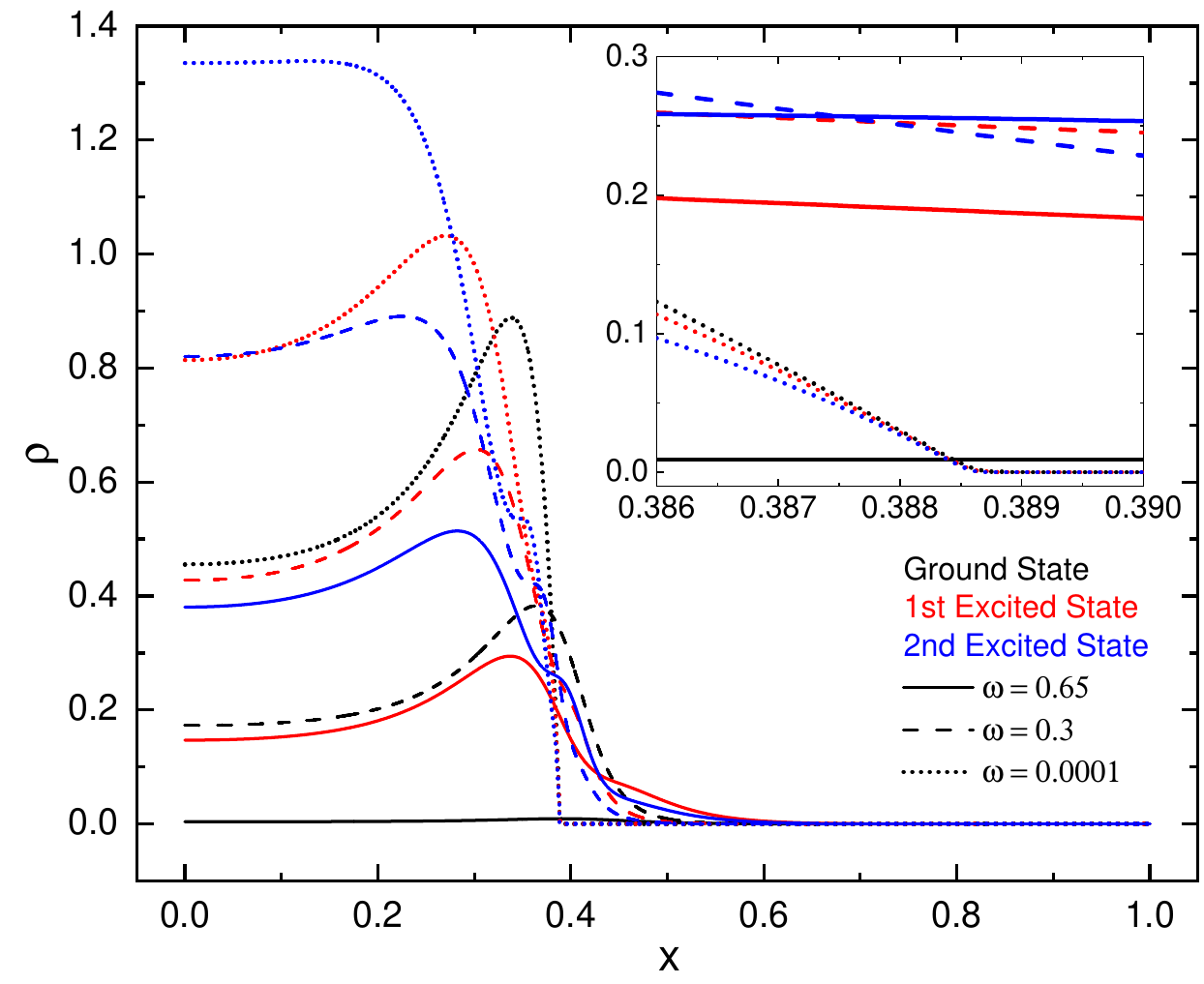}
  \begin{center}
      \includegraphics[width=8.1cm]{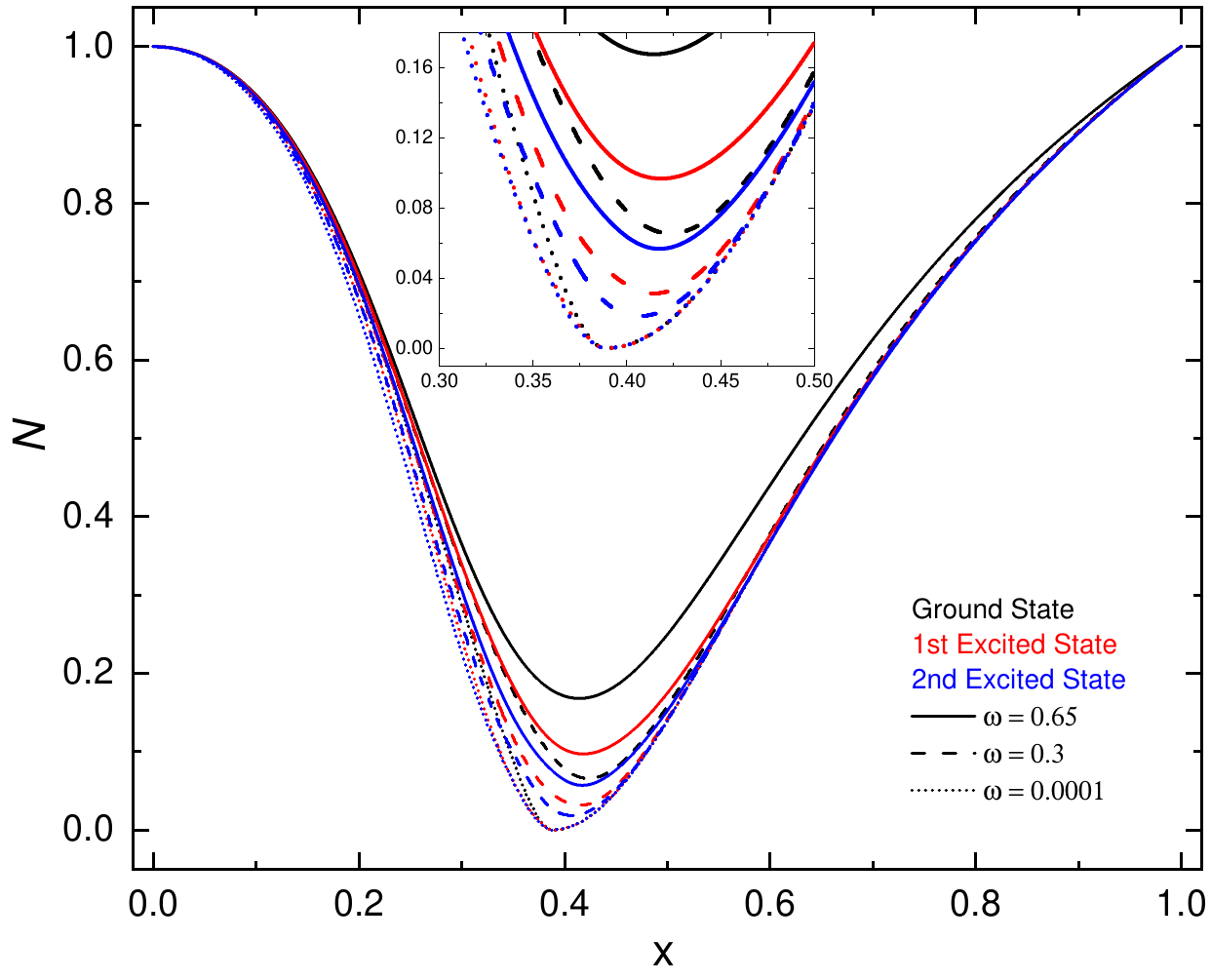}
     \includegraphics[width=8.2cm]{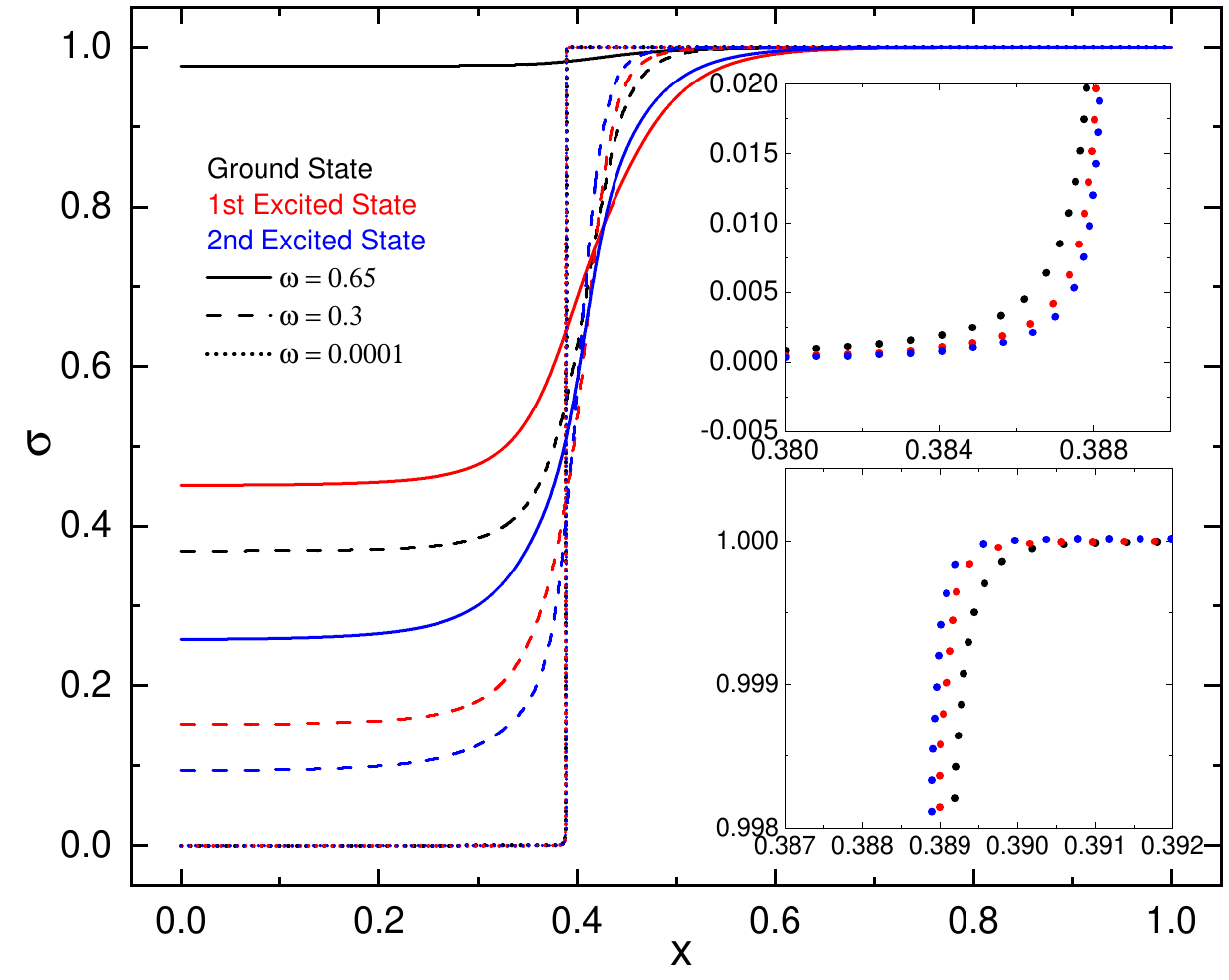}
  \end{center}
  \caption{The radial distribution of the  field functions with $q=0.56$.
  }\label{phase}
\end{figure}

\begin{figure}[]
  \begin{center}
  \includegraphics[width=8.4cm]{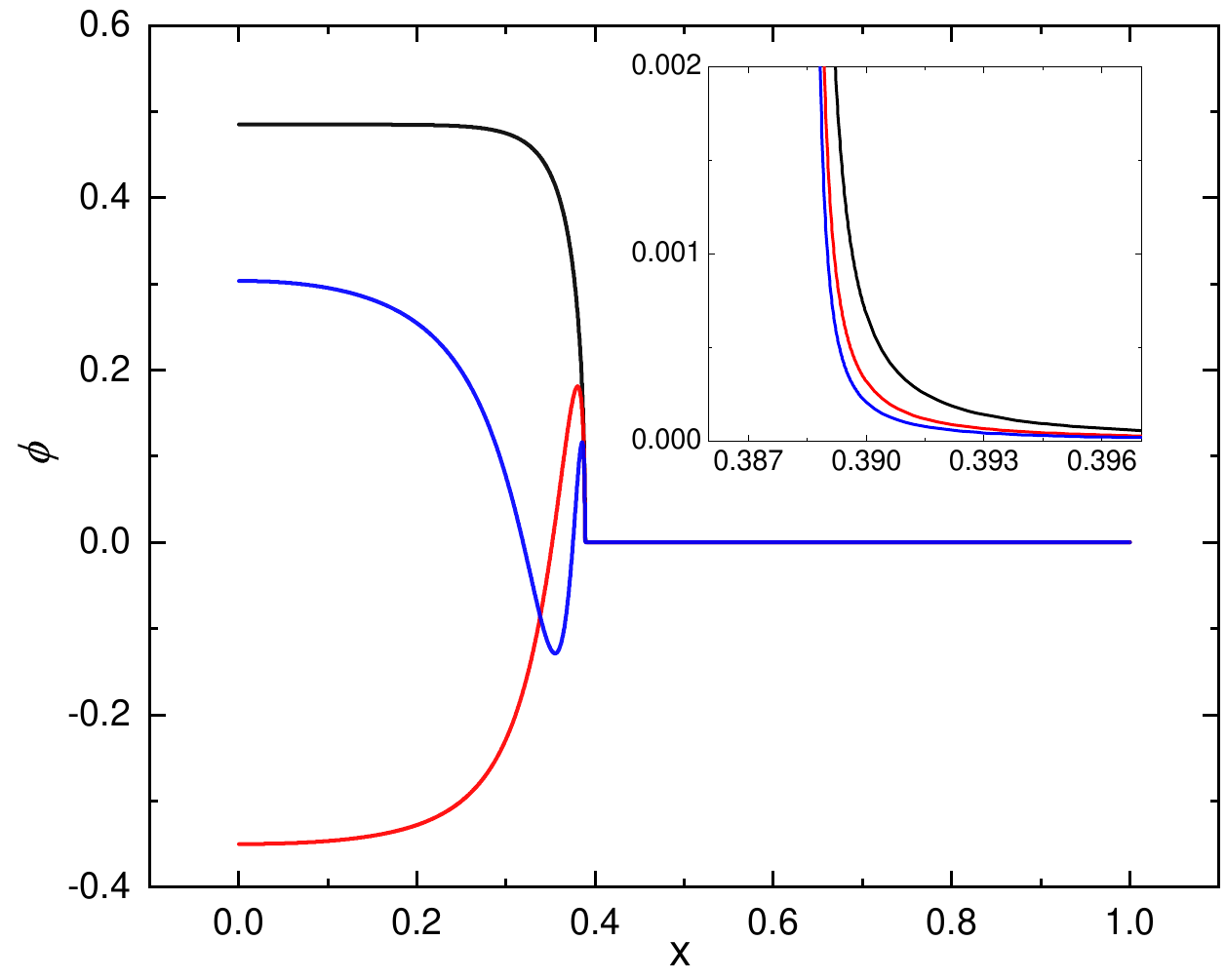}
   \includegraphics[width=7.8cm]{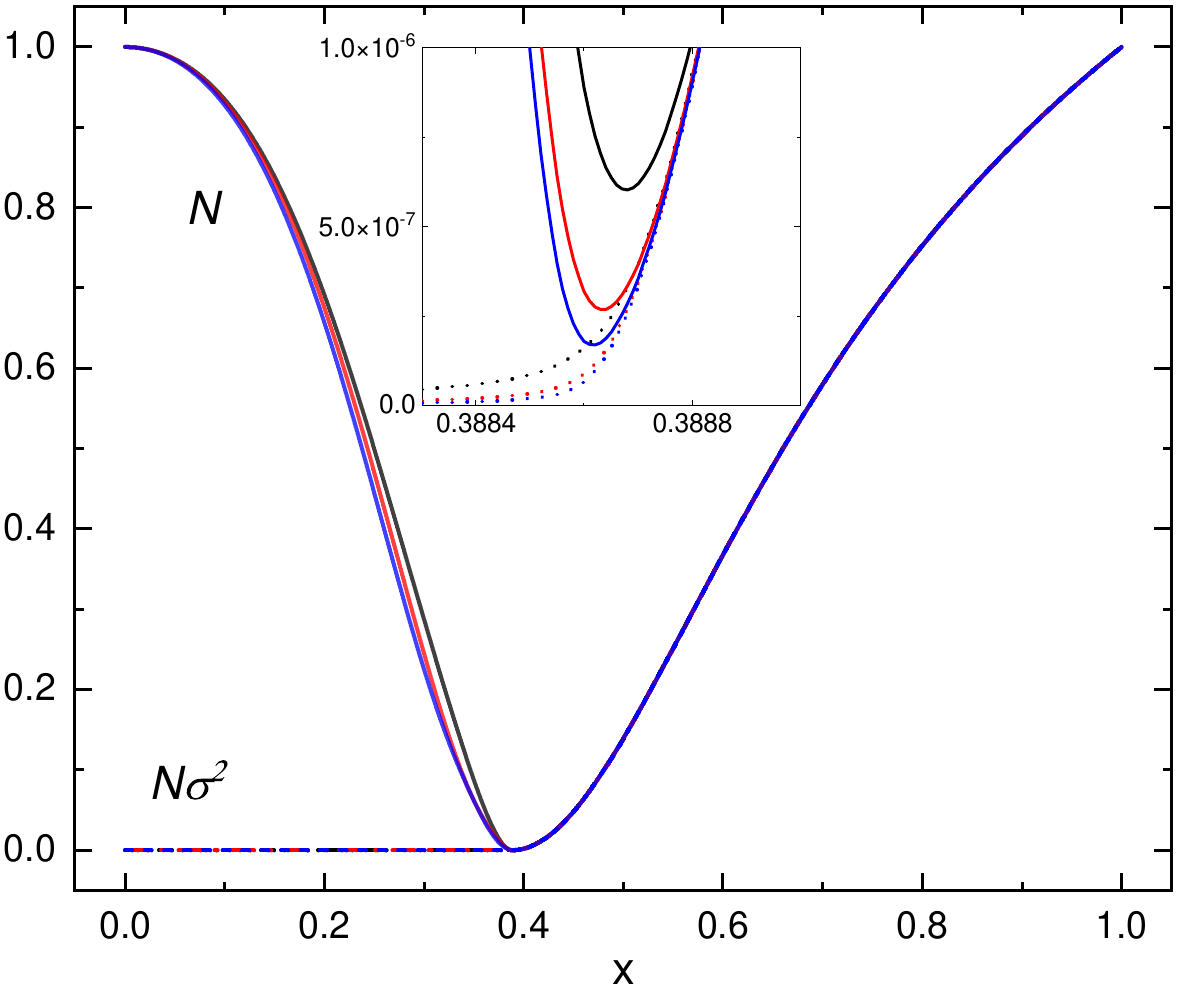}
  \end{center}
  \caption{The radial distribution   of the extreme solutions with $\omega = 0.0001 $.  All curves have the same value of $q=0.56$. And both the ground state and excited states have the same mass $M=0.49612$.
  }\label{phase2}
\end{figure}
 Firstly, in Fig. \ref{phase}, we depict the configuration of Hayward-boson star solutions with varying values of the frequency $\omega$ from ground state to second excited state. 
In both plots, the black, red, and blue lines represent the ground state, first, and second states, respectively. Meanwhile, the solid, dashed, and dotted lines correspond to the frequencies $\omega= 0.65$,  $0.3$, and $0.0001$, respectively.
The insets show the detail of the curves. In the top left plot, we can observe  that the ground state has no nodes, and $n$-th excited state has exactly $n$ nodes,
where the values of the scalar field $\phi$ are equal to zero. Moreover,  at the same frequency, the amplitude of $\phi$ deceases
with higher excited states. However, at the same ground state or excited state, the amplitude increases as the frequency decreases. In addition, the corresponding energy density, denoted as $\rho=-T^{(2)0}_{~~0}$, for the scalar matter field is displayed in the top right plot.
The metric functions $N$ and $\sigma$ are shown in the down  panels. 
At the same frequency, the minimum value of the function $N$ deceases with higher excited states. Meanwhile, at the same ground state or excited state, the minimum value decreases as the frequency decreases. The function $\sigma$ also exhibits the same property.
It is worth noting that as the frequency approaches zero, the scalar field in these solutions converges at nearly the same radial coordinate axis position. The minimum values of 
$N$ and $\sigma$ also rapidly tend toward zero at the same position due to the influence of the matter field. We refer to this position as a critical location, as it is very close to forming the event horizon, and can be termed as  critical horizon.

To provide a more detailed presentation of the scenario when the frequency approaches zero, we show the distribution of   $\phi$, $N$, and  $g_{tt}=N\sigma^2$ as a function of $x$ coordinate with the magnetic charge $q=0.56$ in Fig. \ref{phase2}. From the insets, we can see that the scalar field of these solutions converges 
at critical horizon while rapidly decaying beyond that radius, forming a steep surface. and the $g_{tt}$ is nearly zero inside the critical horizon. This behavior is characteristic of a frozen star.

\begin{figure}[]
  \begin{center}
  \includegraphics[width=8.1cm]{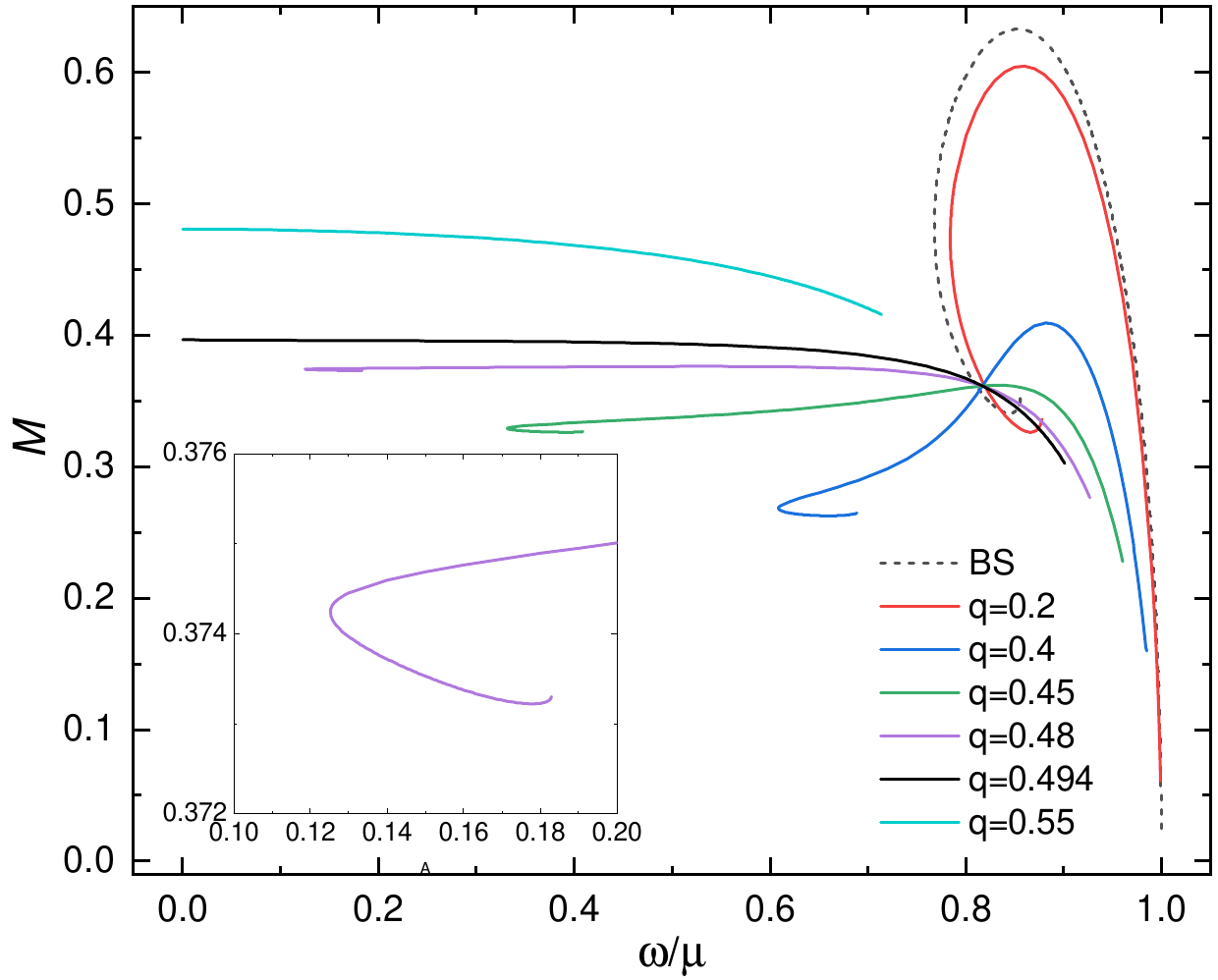}
   \includegraphics[width=8.1cm]{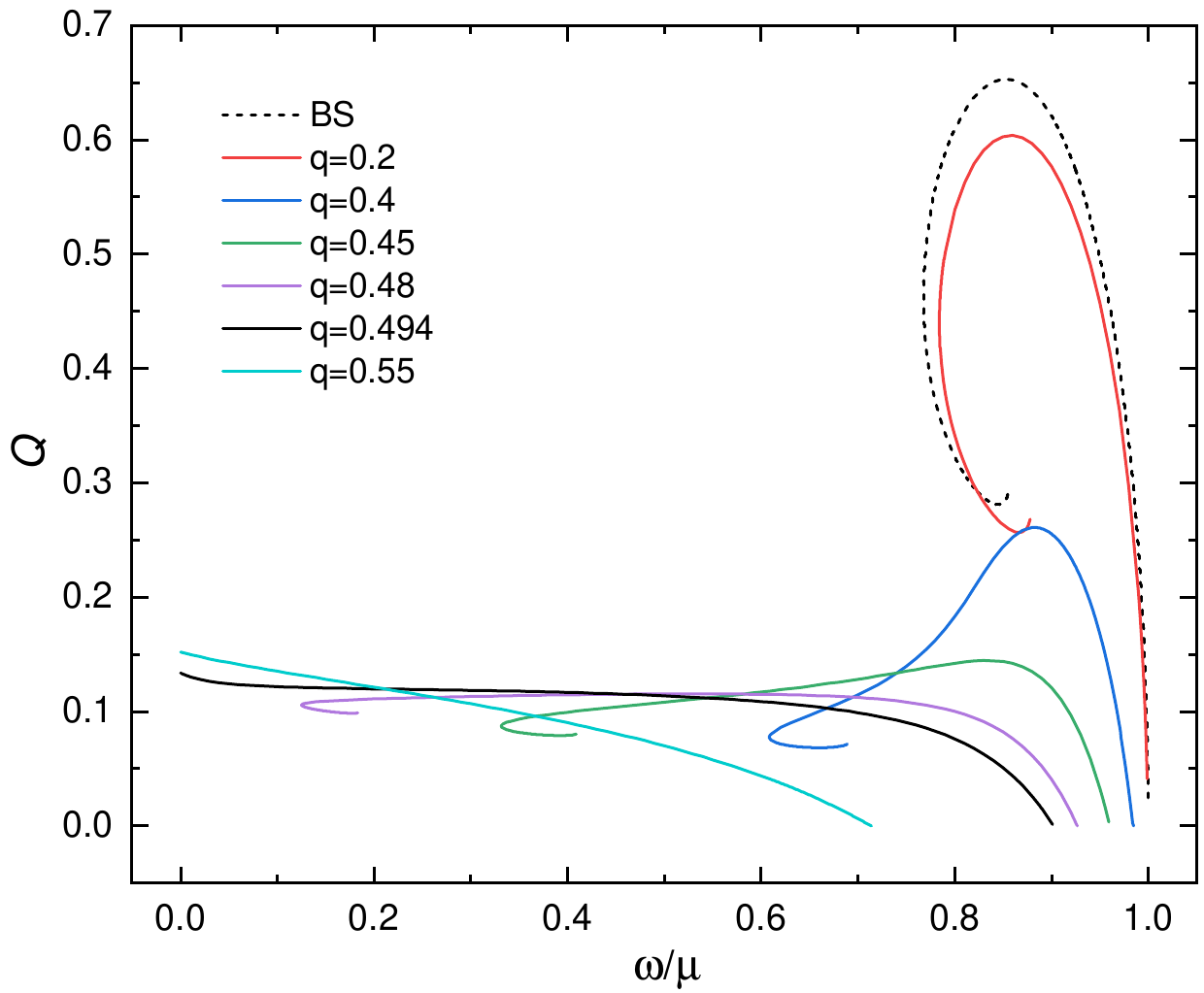}
   \includegraphics[width=8.15cm]{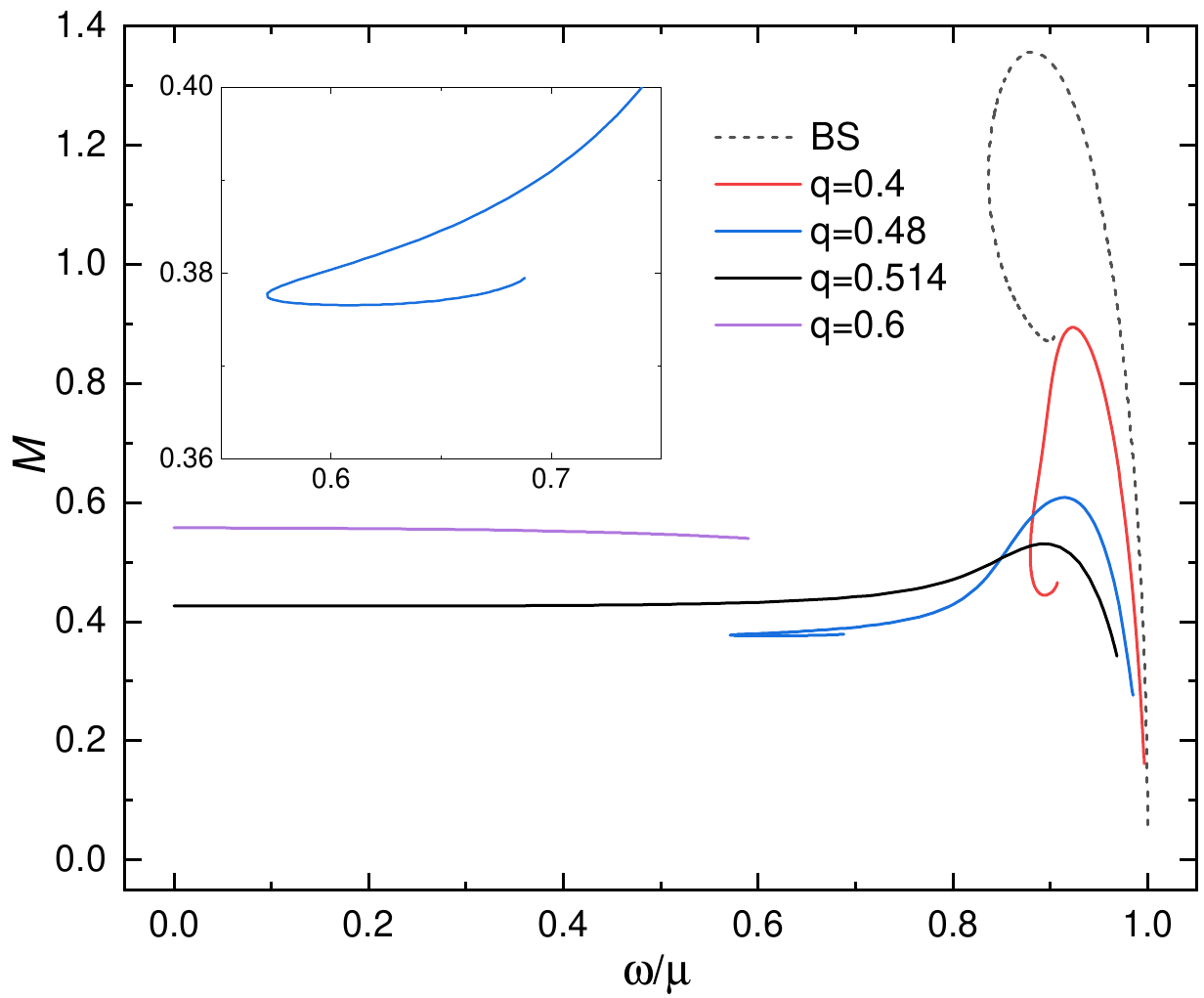}
     \includegraphics[width=8.15cm]{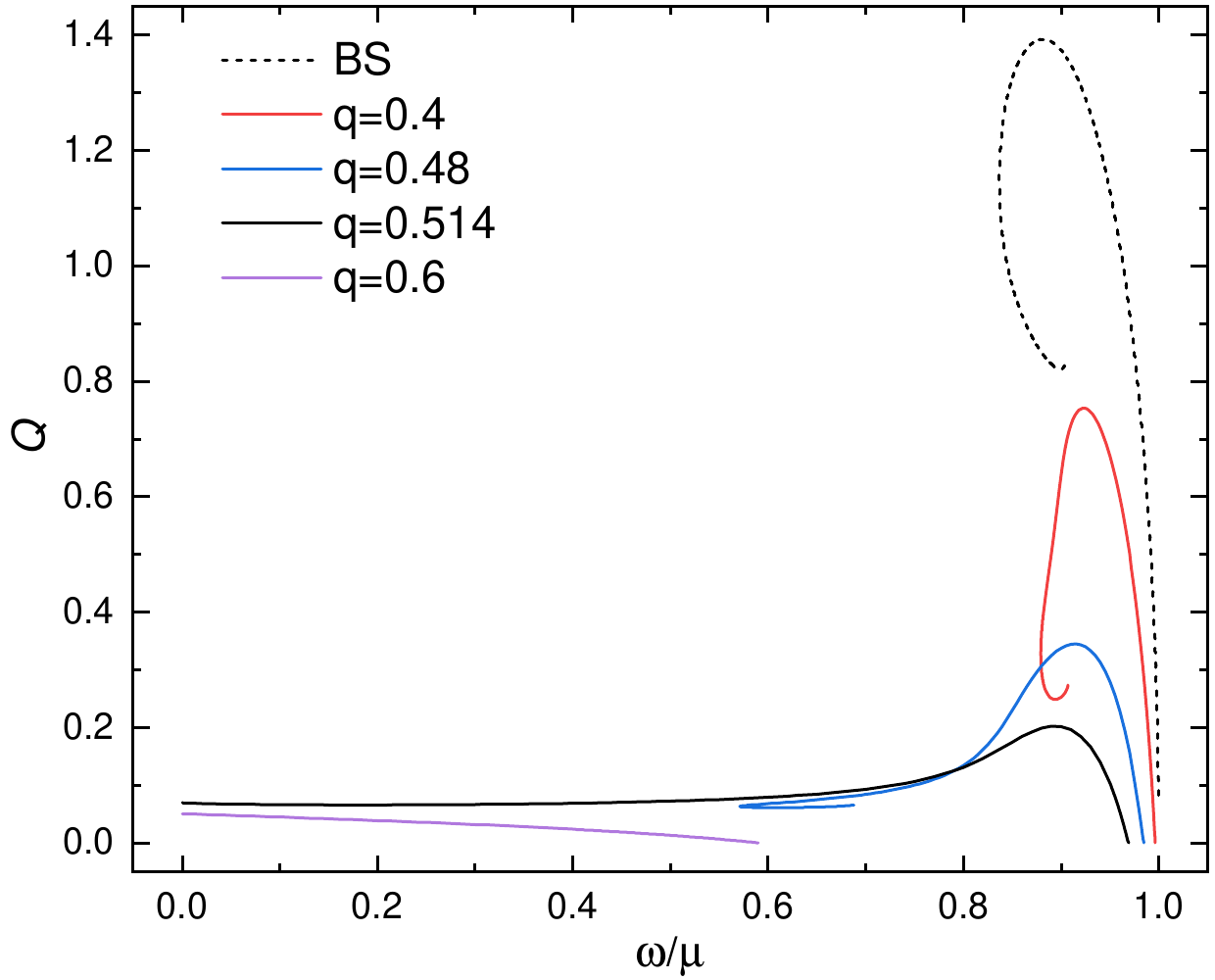}
     \includegraphics[width=8.2cm]{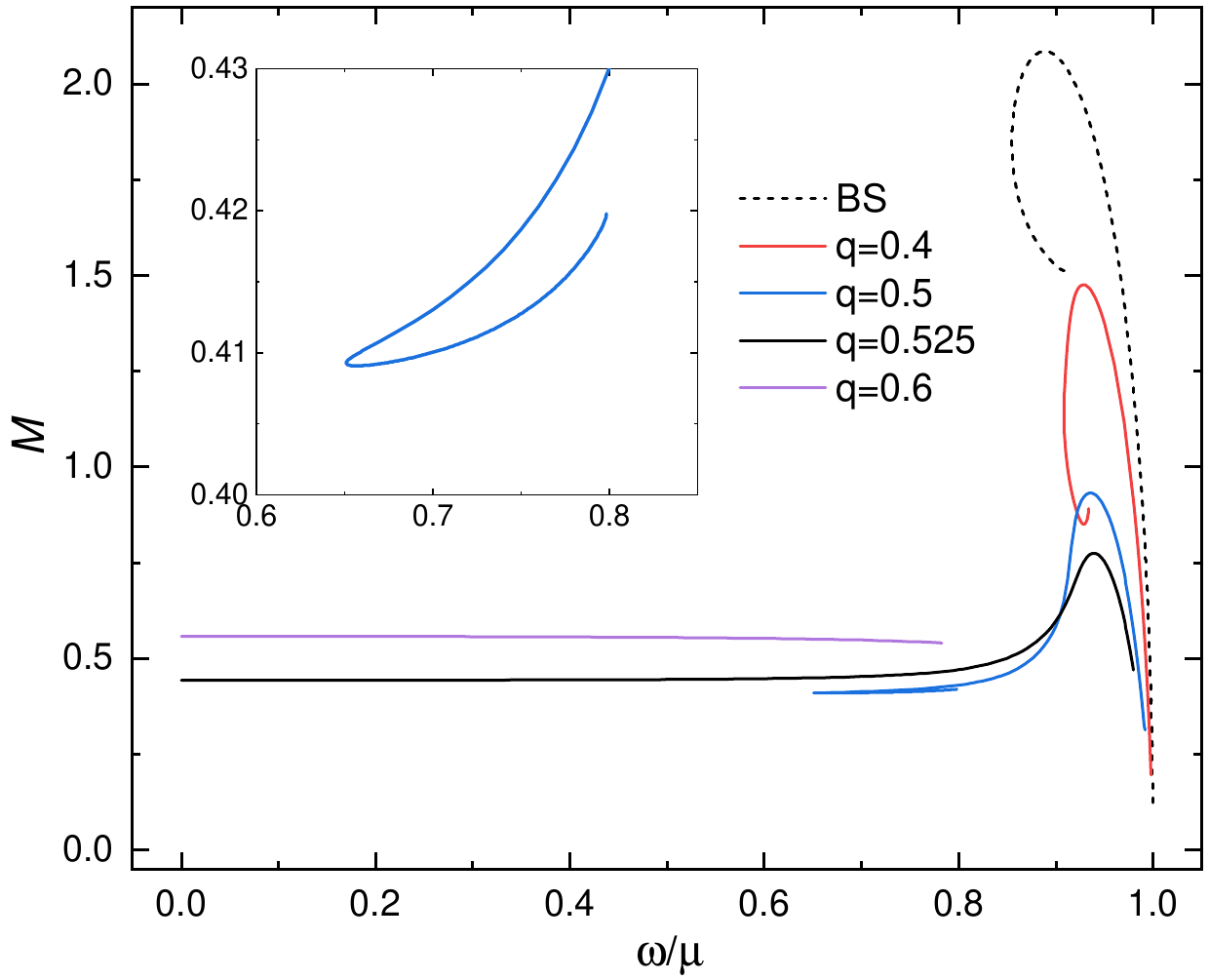}
      \includegraphics[width=8.1cm]{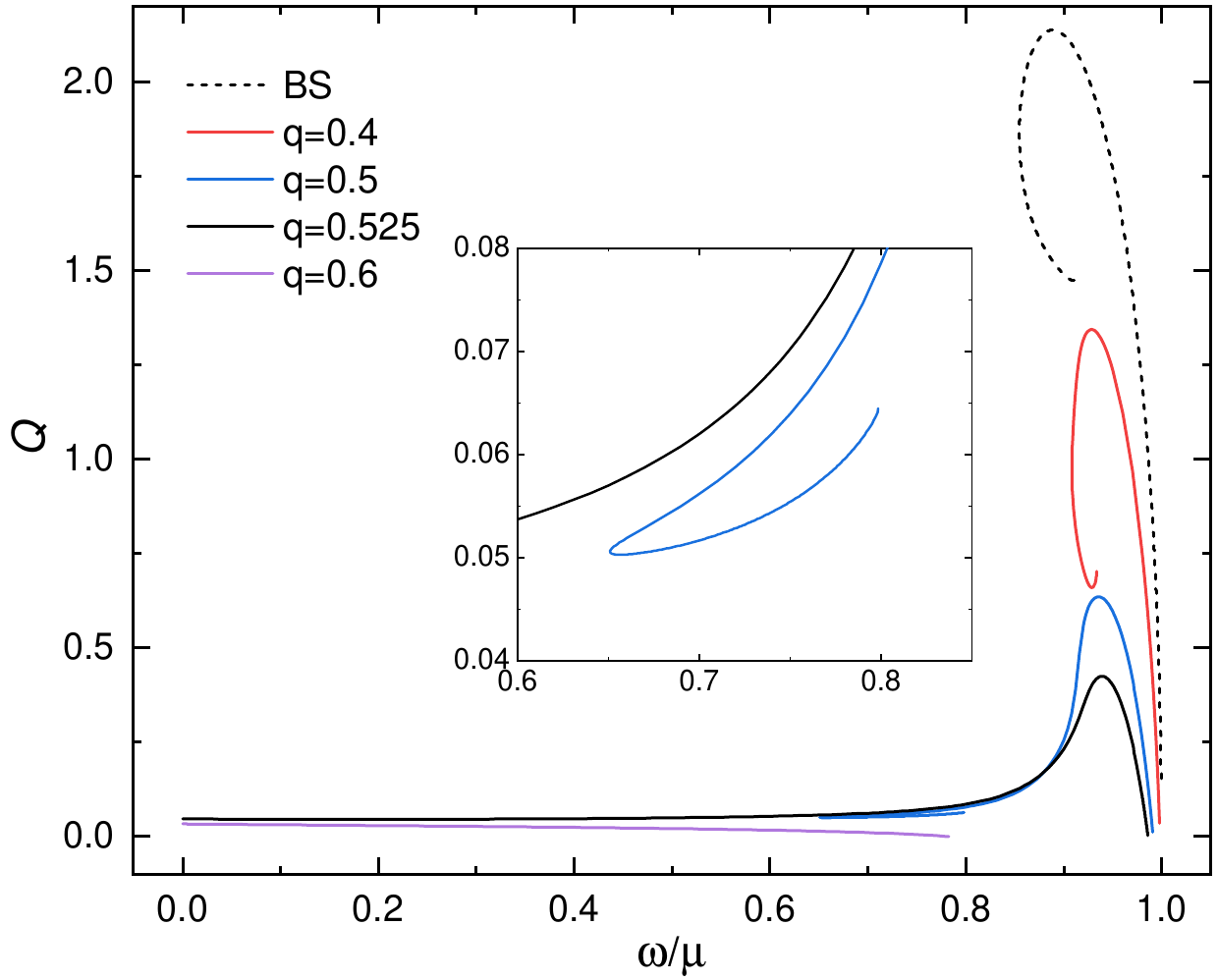}

  \end{center}
  \caption{The mass $M$ and Noether charges $Q$ of complex scalar field versus the frequency $\omega$ with the different values of magnetic charges $q$.
  }\label{phase3}
\end{figure}

In Fig. \ref{phase3}, we
exhibit the mass $M$ and Noether charges $Q$ of  Hayward-boson stars versus the frequency $\omega$ with the different values of magnetic charges $q$  from the ground state (upper panels) to second excited
state  (lower panels).
In all of plots, the black, dotted lines represent the solutions of boson star with $q=0$,  exhibiting a spiral pattern. As the parameter $q$ initiates its ascent from zero, the curves undergo deformation and gradual unfolding. Simultaneously, the extent of the second branch diminishes. Upon surpassing a critical value $q=0.494$ (ground state), 0.514 (first excited state), and 0.525 (second excited state), the second branch vanishes, transforming the multi-valued curve into a single-valued curve, and the minimum frequency can approach zero infinitely, which correspond to frozen stars.

\section{Conclusions}
In this paper, the study of frozen Bardeen-boson  star  is extended to the case of frozen Hayward-boson  star.  We investigated the model of Einstein-Klein-Gordon theory coupled to a nonlinear electrodynamics proposed by S. Hayward. In the absence of the scalar field, the configuration corresponds to the well-established Hayward black hole solution. Nevertheless, in the presence of the scalar field, our attempts to identify a black hole solution proved unsuccessful. Instead, we uncovered a family of  Hayward-boson star solutions, and no event horizon exists. When the magnetic charge $q$ exceeds a certain critical value, the frozen Hayward-boson  star can be obtained with $\omega \rightarrow 0$. In this scenario, the scalar field converges at the critical horizon, experiencing rapid decay beyond this threshold. The critical value of $q$ increase with higher excited states. The energy density of the scalar matter field exhibits a steep, wall-like distribution in the proximity of the inner region near critical horizon. 
From the perspective of an observer at infinity, these frozen star solutions can be regarded as an extremal black hole. Moreover, notably, when subjected to identical parameters, it is noted that both the ground state and the excited states exhibited identical critical horizons and masses.

Several extensions of our study warrant exploration. Firstly, we aim to expand our investigation to formulate rotating generalizations of the frozen Hayward-boson  star. Secondly, considering the study of multistate boson stars in \cite{Bernal:2009zy},  it would be intriguing to construct a frozen multistate Hayward-boson  star, incorporating coexisting states of multiple scalar fields, including both ground and excited states scalar fields. Finally, the stability of these Hayward-boson stars remains unexplored in this paper, and we intend to address this aspect in our subsequent work.

\section{Acknowledgment}
This work is supported by National Key Research and Development Program of China (Grant No. 2020YFC2201503) and the National Natural Science Foundation of China (Grants No.~12275110 and No.~12047501).


\begin{thebibliography}{99}

\bibitem{Penrose:1964wq}
R.~Penrose,
Phys. Rev. Lett. \textbf{14}, 57-59 (1965)
doi:10.1103/PhysRevLett.14.57.
\bibitem{Hawking:1966sx}
S.~Hawking,
Proc. Roy. Soc. Lond. A \textbf{294}, 511-521 (1966)
doi:10.1098/rspa.1966.0221.

\bibitem{Hawking:1970zqf}
S.~W.~Hawking and R.~Penrose,
Proc. Roy. Soc. Lond. A \textbf{314}, 529-548 (1970)
doi:10.1098/rspa.1970.0021.













\bibitem{Einstein:1939ms}
A.~Einstein,
``On a stationary system with spherical symmetry consisting of many gravitating masses,''
Annals Math. \textbf{40}, 922-936 (1939)
doi:10.2307/1968902.







\bibitem{Bardeen1}
J. Bardeen, in Proceedings of GR5, Tiflis, U.S.S.R. (1968).

\bibitem{Ayon-Beato:1998hmi}
E.~Ayon-Beato and A.~Garcia,
``Regular black hole in general relativity coupled to nonlinear electrodynamics,''
Phys. Rev. Lett. \textbf{80}, 5056-5059 (1998)
doi:10.1103/PhysRevLett.80.5056
[arXiv:gr-qc/9911046 [gr-qc]].


\bibitem{Ayon-Beato:2000mjt}
E.~Ayon-Beato and A.~Garcia,
``The Bardeen model as a nonlinear magnetic monopole,''
Phys. Lett. B \textbf{493}, 149-152 (2000)
doi:10.1016/S0370-2693(00)01125-4
[arXiv:gr-qc/0009077 [gr-qc]].

\bibitem{Lan:2023cvz}
C.~Lan, H.~Yang, Y.~Guo and Y.~G.~Miao,
``Regular Black Holes: A Short Topic Review,''
Int. J. Theor. Phys. \textbf{62}, no.9, 202 (2023)
doi:10.1007/s10773-023-05454-1
[arXiv:2303.11696 [gr-qc]].


\bibitem{Wang:2023tdz}
X.~E.~Wang,
``From Bardeen-boson stars to black holes without event horizon,''
[arXiv:2305.19057 [gr-qc]].





\bibitem{Wheeler:1955zz}
J.~A.~Wheeler,  Geons, {\em Phys. Rev.} {\bf 97}, 511 (1955).

\bibitem{Power:1957zz}
E.~A.~Power  and J.~A.~Wheeler, Thermal Geons, {\em Rev. Mod. Phys}. {\bf 29}, 480 (1957).





\bibitem{Kaup:1968zz}
D.~J.~Kaup, Klein-Gordon Geon, {\em Phys. Rev}. {\bf 172}, 1331 (1968).

\bibitem{PhysRev.187.1767}
 R.~Ruffini and S.~Bonazzola, Systems of Self-Gravitating Particles in
  General Relativity and the Concept of an Equation of State, {\em Phys.
  Rev}. {\bf 187}, 1767(1969).




\bibitem{Schunck:2003kk}
F.~E.~Schunck and E.~W.~Mielke,
``General relativistic boson stars,''
Class. Quant. Grav. \textbf{20}, R301-R356 (2003)
doi:10.1088/0264-9381/20/20/201
[arXiv:0801.0307 [astro-ph]].



\bibitem{Liebling:2012fv}
S.~L.~Liebling and C.~Palenzuela,
``Dynamical boson stars,''
Living Rev. Rel. \textbf{15}, 6 (2012)
doi:10.1007/s41114-023-00043-4
[arXiv:1202.5809 [gr-qc]].

\bibitem{Oppenheimer:1939ue}
J.~R.~Oppenheimer and H.~Snyder,
``On Continued gravitational contraction,''
Phys. Rev. \textbf{56}, 455-459 (1939)
doi:10.1103/PhysRev.56.455.


\bibitem{zeldovichbookorpaper} Y. B. Zel'dovich and I. D. Novikov,
{\it Relativistic Astrophysics 1: Stars and Relativity}, (University
of Chicago Press, Chicago 1971), p. 369 (translation from the 1967
Russian edition).

\bibitem{Ruffini:1971bza}
R.~Ruffini and J.~A.~Wheeler,
``Introducing the black hole,''
Phys. Today \textbf{24} (1971) no.1, 30

\bibitem{Hayward:2005gi}
S.~A.~Hayward,
``Formation and evaporation of regular black holes,''
Phys. Rev. Lett. \textbf{96}, 031103 (2006)
doi:10.1103/PhysRevLett.96.031103
[arXiv:gr-qc/0506126 [gr-qc]].



\bibitem{Trefethen}
Trefethen, L.N. 
 {{\em Spectral Methods in MATLAB}; SIAM: Philadelphia, PA, USA, 2000.}


\bibitem{Bernal:2009zy}
A.~Bernal, J.~Barranco, D.~Alic and C.~Palenzuela,
``Multi-state Boson Stars,''
Phys. Rev. D \textbf{81}, 044031 (2010)
doi:10.1103/PhysRevD.81.044031
[arXiv:0908.2435 [gr-qc]].



\end{thebibliography}
\end{document}